\documentclass[aps,prc,showpacs,onecolumn,superscriptaddress]{revtex4}
\usepackage{CJK}
\usepackage{multirow}
\usepackage{graphicx}
\usepackage{color}
\usepackage{amsmath}
\usepackage{amssymb}
\usepackage[colorlinks=true]{hyperref}

\definecolor{dgreen}{cmyk}{1.,0.,1.,0.4}        
\definecolor{orange}{cmyk}{0.,0.353,1.,0.}    

\begin{document}
\begin{CJK*}{GB}{} 
\title{
\begin{flushright}
{
}
\end{flushright}
Review of anisotropic flow correlations in ultrarelativistic heavy-ion collisions }

\author{You~Zhou}
\email{you.zhou@cern.ch}
\affiliation{Niels Bohr Institute, University of Copenhagen, Blegdamsvej 17, 2100 Copenhagen, Denmark}

\date{\today}

\begin{abstract}

Anisotropic flow phenomena is a key probe of the existence of Quark-Gluon Plasma. Several new observable associated with correlations between anisotropic flow harmonics are developed, which are expected to be sensitive to the initial fluctuations and transport properties of the created matter in heavy ion collisions. 
I review recent developments of correlations of anisotropic flow harmonics. The experimental measurements, together with the comparisons to theoretical model calculations, open up new opportunities of exploring novel QCD dynamics in heavy-ion collisions.

\end{abstract}



\maketitle
\end{CJK*}


\section{Introduction}
\label{section1}

One of the fundamental questions in the phenomenology of Quantum Chromo Dynamics (QCD) 
is what are the properties of matter at extreme densities and temperatures where quarks
and gluons are in a new state of matter, the so-called Quark Gluon Plasma (QGP)~\cite{Lee:1978mf, Shuryak:1980tp}.
Collisions of high-energy heavy-ions, at the Brookhaven Relativistic Heavy Ion Collider (RHIC) 
and the CERN Large Hadron Collider (LHC), allow us to create and study the properties of the QGP matter in the laboratory.
This matter expands under large pressure gradients, which transfer the inhomogeneous initial conditions into azimuthal anisotropy of produced particles in momentum space. 
This anisotropy of produced particles is one of the probes of the properties of the QGP~\cite{Ollitrault:1992bk, Voloshin:1994mz}. It can be characterized by an expansion of the single-particle azimuthal distribution $P(\varphi)$: 
\begin{equation}
P(\varphi) = \frac{1}{2\pi} \sum_{n=-\infty}^{+\infty} {\overrightarrow{V_{n}} \, e^{-in\varphi} }
\end{equation}
where $\varphi$ is the azimuthal angle of emitted particles, $\overrightarrow{V_{n}}$ is the $n$-th order flow-vector defined as $\overrightarrow{V_{n}} = v_{n}\,e^{in\Psi_{n}}$, its magnitude $v_{n}$ is the $n$-th order anisotropic flow harmonic and its orientation is symmetry plane (participant plane) angle $\Psi_{n}$. Alternatively, this anisotropy can be generally given by the joint probability density function ($p.d.f.$) in terms of $v_{n}$ and $\Psi_{n}$ as:
\begin{equation}
P(v_{m}, v_{n}, ..., \Psi_{m}, \Psi_{n}, ...) = \frac{1}{N_{event}} \frac{dN_{event}} {v_{m} v_{n} \cdot \cdot \cdot {\rm d} v_{m} \,{\rm d} v_{n} \cdot \cdot \cdot {\rm d}\Psi_{m} \, {\rm d}\Psi_{n} \cdot \cdot \cdot}
\end{equation}
In the last decade, the experimental measurements of anisotropic flow $v_{n}$~\cite{Alt:2003ab, Ackermann:2000tr, Adler:2001nb, Adler:2002pu, Adams:2004bi, Adams:2005zg,  Abelev:2008ae, Adamczyk:2013gw, Adcox:2002ms, Adler:2003kt, Adare:2006ti, Afanasiev:2009wq, Adare:2010ux, Adare:2011tg, Back:2002gz, Back:2004mh, Back:2004zg, Manly:2005zy, Back:2005pc, Alver:2006wh, Alver:2007qw, Alver:2010rt,
Aamodt:2010pa, ALICE:2011ab, Abelev:2012di, Abelev:2013cva, Abelev:2014mda, Abelev:2014pua, Adam:2015eta, Adam:2015vje, Adam:2016izf, ALICE:2016kpq, ATLAS:2011ah, ATLAS:2012at, Aad:2013fja, Aad:2013xma, Aad:2014fla, Aad:2014eoa, Aad:2014vba, Aad:2014lta, Aad:2015lwa, 
Chatrchyan:2012wg, Chatrchyan:2012xq, Chatrchyan:2012ta, Chatrchyan:2012vqa,  Chatrchyan:2013nka, Chatrchyan:2013kba,  CMS:2013bza, Khachatryan:2014jra, Khachatryan:2015waa, Khachatryan:2015oea}, combined with theoretical advances from calculations made in a variety of frameworks~\cite{Huovinen:2001cy, Kolb:2000fha, Luzum:2008cw, Song:2007ux, Song:2010mg, Niemi:2015qia}, have led to a broad and deep knowledge of initial conditions and properties of the created hot/dense QCD matter.
In particular, the precision anisotropic flow measurements based on the huge data collected at the LHC experiments and the successful description from hydrodynamic calculations demonstrate that the QGP created in heavy ion collisions behaves like a strongly coupled liquid with a very small specific shear viscosity $\eta/s$~\cite{Heinz:2013th, Luzum:2013yya, Huovinen:2013wma, Shuryak:2014zxa, Song:2013gia, Dusling:2015gta}, which is close to a quantum limit 1/4$\pi$~\cite{Kovtun:2004de}.

It has been investigated into great details of event-by-event fluctuations of single flow harmonic. Based on the measurements of higher order cumulants of anisotropic flow~\cite{Collaboration:2011yba, Aad:2014vba, Chatrchyan:2012ta, Chatrchyan:2013kba, Zhou:2016fvj} and the event-by-event $v_{n}$ distributions~\cite{Aad:2013xma}, it was realized that the newly proposed Elliptic-Power function~\cite{Yan:2014afa,Yan:2013laa,Yan:2014nsa} gives the best description of underlying $p.d.f.$ of single harmonic $v_{n}$ distributions~\cite{ Bravina:2015sda, Zhou:2015eya, Jia:2014jca}. On the other hand, it has been known for a while that both the flow harmonic (magnitude) $v_{n}$ and its symmetry plane (orientation) $\Psi_{n}$ of the flow-vector  $\overrightarrow{V_{n}}$ fluctuate event-by-event~\cite{Petersen:2010cw, Qiu:2011iv, Niemi:2012aj}, but only recently the $p_{\rm T}$ and $\eta$ dependent flow angle ($\Psi_{n}$) and magnitude ($v_{n}$) were predicted by hydrodynamic calculations~\cite{Heinz:2013bua, Gardim:2012im}. Many indications were quickly obtained in experiments by looking at the deviations from unity of $v_{n}[2]/v_{n}\{2\}$~\cite{Zhou:2014bba} and factorization ratio $r_{n}$~\cite{Zhou:2014bba, CMS:2013bza, Khachatryan:2015oea}. These measurements were nicely predicted or reproduced by hydrodynamic calculations, and are found to be sensitive to either the initial-state density fluctuations and/or the shear viscosity of the expanding fireball medium~\cite{Heinz:2013bua, Gardim:2012im, Kozlov:2014hya}. Most of these above mentioned studies are focused on the fluctuations of single flow harmonics and its corresponding symmetry planes, as a function of collisions centrality, transverse momentum $p_{\rm T}$ and pseudorapidity $\eta$. 
Results of correlations between symmetry planes~\cite{Aad:2014fla, ALICE:2011ab} reveal a new type of correlations between different order flow-vectors, which was investigated in the observable of $v_{2n/\Psi_{n}}$ before~\cite{Andronic:2000cx, Chung:2001qr, Adams:2003zg}. In particular, some of the symmetry planes correlations show quite different centrality dependence from the initial- and final-state, and this characteristic sign change during system evolution is correctly reproduced by theoretical calculations~\cite{Qiu:2011iv, Teaney:2013dta, Niemi:2015qia}, thus confirms the validity of hydrodynamic framework in heavy-ion collisions and further yields valuable additional insights into the fluctuating initial conditions and hydrodynamic response~\cite{Qiu:2011iv, Bhalerao:2013ina, Niemi:2015qia}.

In addition to all these observables, the (anti-)correlations between anisotropic flow harmonics $v_{m}$ and $v_{n}$ are found to be extremely interesting~\cite{Bilandzic:2013kga, Bhalerao:2014xra, Niemi:2015qia, Aad:2015lwa, SC:hydro}. A completely new set of information on the joint $p.d.f.$ is carried by the rich pattern observed in experiments. On the other hand, no existed theoretical calculations~\cite{Bilandzic:2013kga, Bhalerao:2014xra, Niemi:2015qia, SC:hydro} could provide quantitative descriptions of data~\cite{ALICE:2016kpq}. Thus, it's crucial to investigate in depth of the relationship between different flow harmonics: whether they are correlated, anti-correlated or not correlated, from both experimental and theoretical point of view.

%

\section{Correlations of $v_{n}$ and $v_{m}$ fluctuations}
\label{sec:vmvnCorrelations}

It is found recently that the relationship between different order flow harmonics can be used to probe the initial-state conditions and the hydrodynamic response of the QGP~\cite{Bilandzic:2013kga, Bhalerao:2011yg, Teaney:2012ke, Bhalerao:2014xra, ALICE:2016kpq}. 
In order to better understand the event-by-event $P(\varphi)$ distribution, it's critical to investigate the relationship between $v_{m}$ and $v_{n}$. 
Considering the naive ellipsoidal shape of the overlap region in non-central heavy ion collisions generates non-vanishing even flow harmonics $v_{2n}$, the correlations between the even flow harmonics are expected. However, it is not straightforward to use geometrical argument to explain the relationship between even flow harmonics for central collisions, where all the harmonics are driven by fluctuations instead of geometry, and to explain the relationship between even and odd odd flow harmonics for central and non-central collisions~\cite{Jia:2014jca}.
A linear correlation function $c(v_{m},v_{n})$ was proposed to study the relationship between $v_{m}$ and $v_{n}$~\cite{Niemi:2012aj}. It is defined as:
\begin{equation}
c(v_{m}, v_{n}) = \left<  \frac{ (v_{m} - \left< v_{m}\right>_{ev} ) \,  (v_{n} - \left< v_{n}\right>_{ev}  )} {\sigma_{v_{m}} \, \sigma_{v_{n}} } \right>_{ev},
\end{equation}
where $\sigma_{v_{m}}$ is the standard deviation of the quantity $v_{m}$, $c(v_{m}, v_{n})$ is 1 (or -1) if $v_{m}$ and $v_{n}$ are linearly (anti-linearly) correlated, and is 0 if not correlated.
\begin{figure}[h]
\centering
\includegraphics[width=0.45\textwidth]{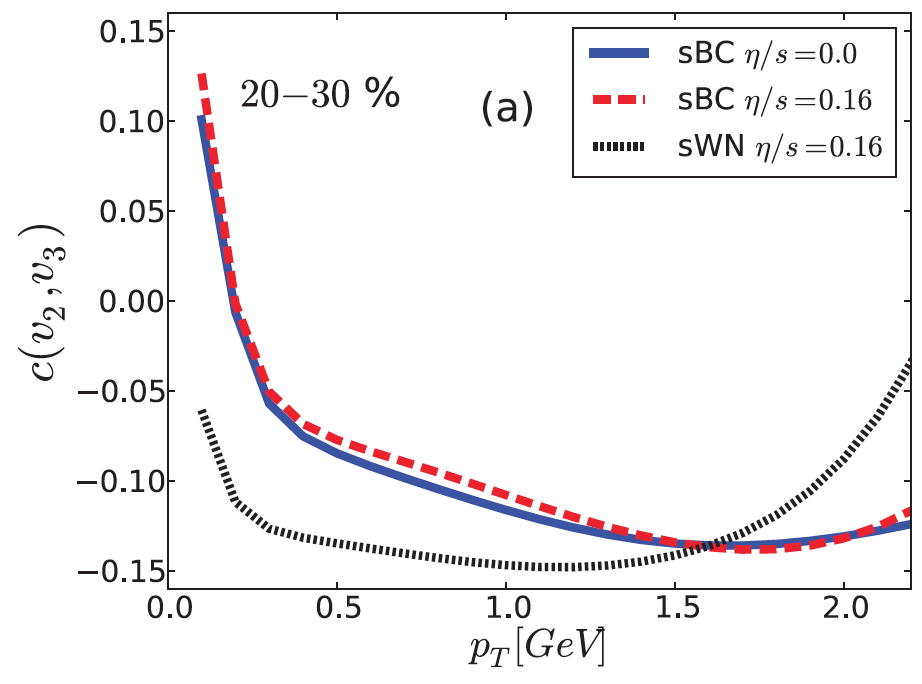}
\includegraphics[width=0.42\textwidth]{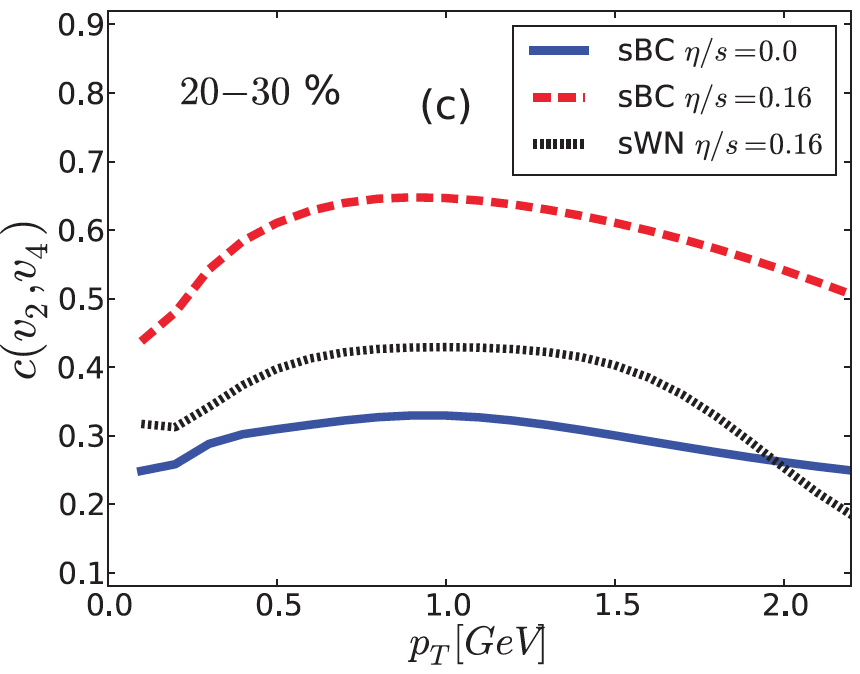}
\caption{(Color online) The $p_{\rm T}$ dependence of $c(v_{2},v_{3})$ (left) and $c(v_{2},v_{4})$ (left) in centrality 20-30\% in Pb--Pb collisions at $\sqrt{s_{_{\rm NN}}}$ = 2.76 TeV. Figures taken from Ref.~\cite{Niemi:2012aj}.}
\label{fig:C23C24}
\end{figure}
It was shown in Fig~\ref{fig:C23C24} that there is an anti-correlationsbetween $v_{2}$ and $v_{3}$, while a correlation was observed between $v_{2}$ and $v_{4}$. In addition, it was demonstrated that $c(v_{2},v_{4})$ depends on both the initial conditions and $\eta/s$ while $c(v_{2},v_{3})$ is only sensitive to $\eta/s$~\cite{Niemi:2012aj}. 
Nevertheless, it cannot be accessible easily in experimental measurements, which rely on two- and multi-particle correlations techniques. 
Thus, it is critical to find an observable which studies the relationship between flow harmonics without contributions from symmetry plane correlations, and can be accessed with observable techniques from experiments. Two different approaches, named ${\it Event\,Shape\,Engineering}$ and ${\it Symmetric\,Cumulant}$, are discussed in the following section.

\subsection{Event Shape Engineering (ESE)}

\begin{figure}[h]
\centering
\includegraphics[width=0.9\textwidth]{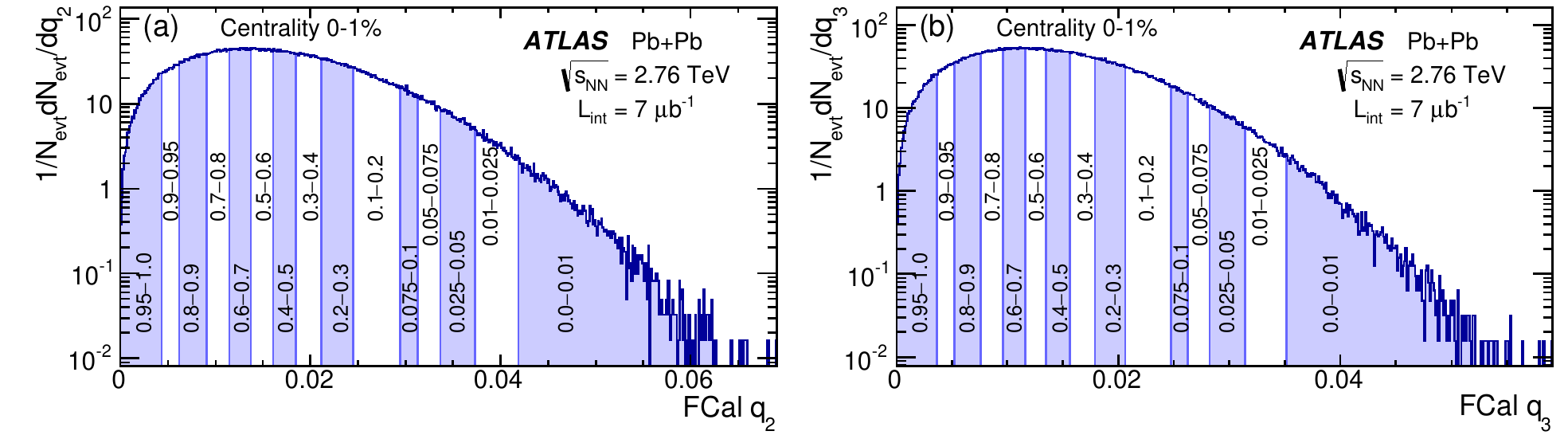}
\caption{(Color online) Distributions of $V_{2}$ (left) and $V_{3}$ (right) calculated with ATLAS forward calorimeter for centrality interval 0-1 \%. Figures taken from Ref.~\cite{Aad:2015lwa}.}
\label{fig:q2q3dis}
\end{figure}

\begin{figure}[h]
\centering
\includegraphics[width=0.8\textwidth]{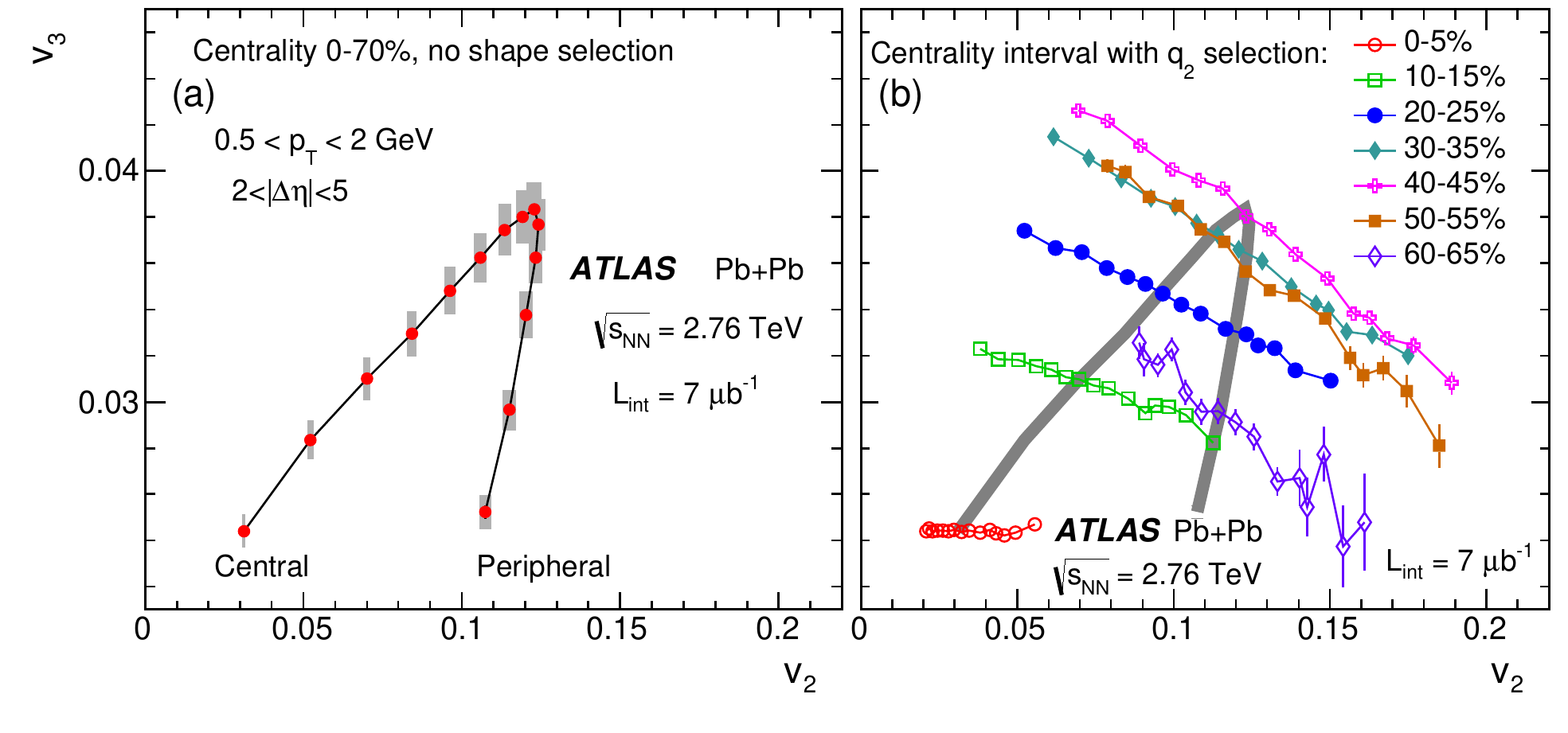}
\caption{(Color online) The correlation of $v_{2}$ (x axis) with $v_{3}$ (y axis) measured in 0.5 $< p_{\rm T} <$ 2 GeV/$c$. The left panel shows the $v_2$ and $v_3$ values for fourteen 5\% centrality intervals over the centrality range 0-70\% without event-shape selection. 
The right panel shows the $v_{2}$ and $v_{3}$ values in the 15 $q_2$ intervals in seven centrality ranges (markers) with larger $v_2$ value corresponding to larger $q_2$ value. Figures taken from Ref.~\cite{Aad:2015lwa}.}
\label{fig:ESE_v23}
\end{figure}

The first experimental attempt was made by ATLAS Collaboration~\cite{Aad:2015lwa}, using the Event-Shape Engineering (ESE)~\cite{Schukraft:2012ah}. This is a technique to select events according to the magnitude of reduced flow vector $\overrightarrow{V_{n}}$. 
Fig.~\ref{fig:q2q3dis} shows the performance of event shape selection on $V_{2}$ (left) and $V_{3}$ (right) in ATLAS detector. For each centrality the data sample is divided into several event classes according to the $V_{2}$ or $V_{3}$ distributions.
Then the $v_{2}$ and $v_{3}$ relationship was investigated by measurements of $v_{2}$ and $v_{3}$ in each event class from ESE selection.
Without using ESE selection, a boomerang-like patten was observed for the centrality dependence of $v_{2}$-$v_{3}$ correlation. This is mainly due to the fact that $v_{3}$ has a weaker centrality dependence than $v_{2}$.
By using ESE, it was observed in Fig.~\ref{fig:ESE_v23} (right) that for event class with the same centrality (shown as the same color), $v_{3}$ decreases as $v_{2}$ increasing. It suggests that $v_{2}$ is anti-correlated with $v_{3}$. 
Considering the linear hydrodynamic response of $v_{2}$ and $v_{3}$ from eccentricity $\varepsilon_{2}$ and triangularity $\varepsilon_{3}$, the anti-correlation between $v_{2}$ and $v_{3}$ might reveal the anti-correlation between $\varepsilon_{2}$ and $\varepsilon_{3}$ of the initial geometry. This indication of initial anti-correlations between $\varepsilon_{2}$ and $\varepsilon_{3}$ is observed in model calculations~\cite{Schukraft:2012ah, Huo:2013qma}.

\begin{figure}[h]
\centering
\includegraphics[width=0.8\textwidth]{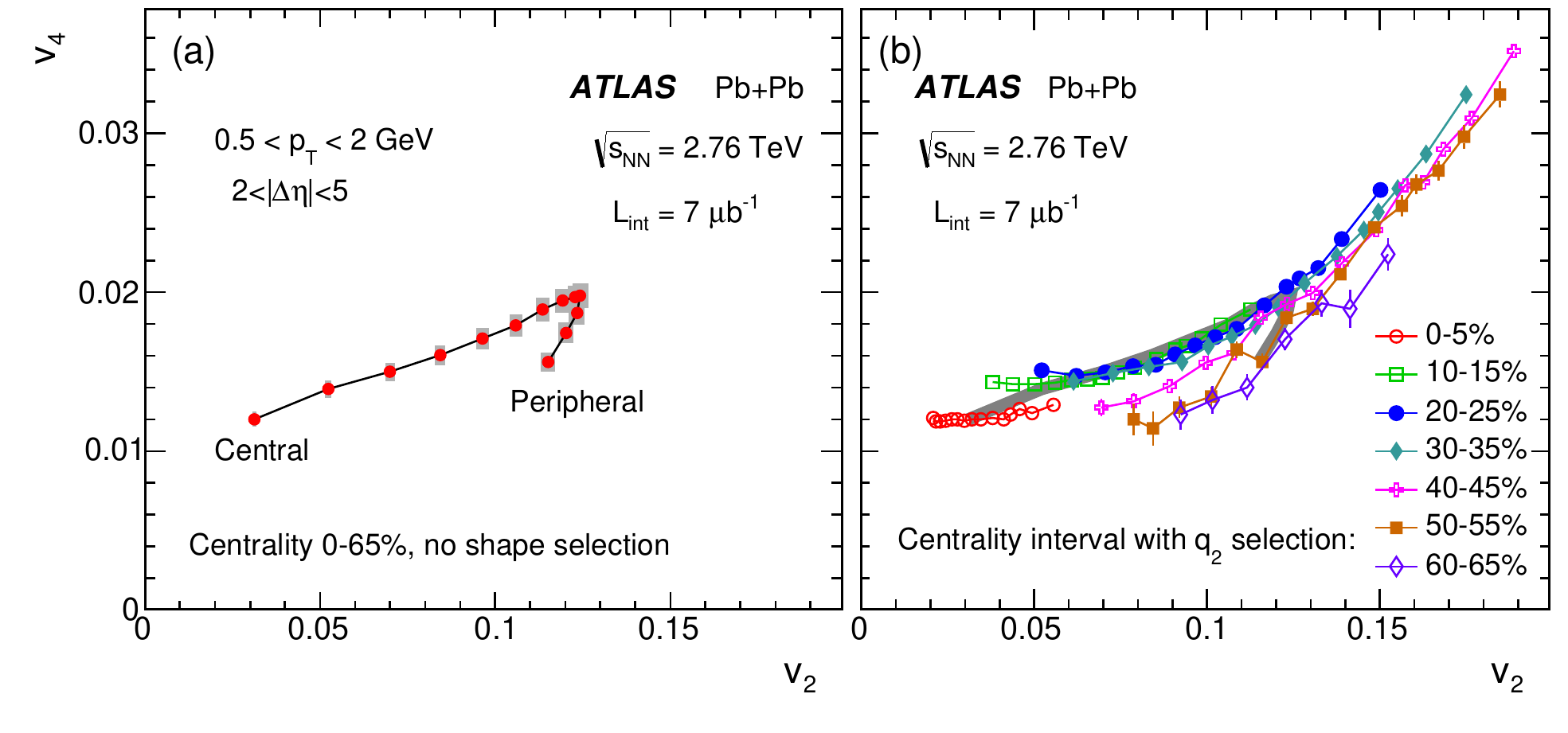}
\caption{(Color online) The correlation of $v_{2}$ (x axis) with $v_{4}$ (y axis) measured in 0.5 $< p_{\rm T} <$ 2 GeV/$c$. The left panel shows the $v_2$ and $v_3$ values for fourteen 5\% centrality intervals over the centrality range 0-70\% without event-shape selection. 
The right panel shows the $v_{2}$ and $v_{4}$ values in the 15 $q_2$ intervals in seven centrality ranges (markers) with larger $v_2$ value corresponding to larger $q_2$ value. Figures taken from Ref.~\cite{Aad:2015lwa}.}
\label{fig:ESE_v24}
\end{figure}

Figure~\ref{fig:ESE_v24} shows the investigation of relationship between $v_{2}$ and $v_{4}$. 
A boomerang-like patten, although weaker than that for the $v_{2}$-$v_{3}$ relationship shown in Fig.~\ref{fig:ESE_v23} (left), is observed in Fig.~\ref{fig:ESE_v24} (left), prior to the ESE selection.
After the ESE selection, it is found in Fig.~\ref{fig:ESE_v24} (right) that $v_{4}$ increases with increasing $v_{2}$. This suggests a correlation between the two harmonics and it can be understood by the interplay between linear and nonlinear collective dynamics in the system evolution~\cite{Aad:2015lwa}. This non-linear contribution of $v_{4}$ from $v_{2}$ is further investigated by fitting the correlation pattern using $v_{4} = \sqrt{c_{0}^{2} + (c_{1} v_{2}^2)^2}$, where $c_{0}$ and $c_{1}$ denote the linear and non-linear components. It is found that the linear component has a weak centrality dependence, while the non-linear component, increasing dramatically with collision centrality, becomes the dominant contribution in the most peripheral collisions~\cite{Aad:2015lwa}.

These (anti)correlation patten between $v_{m}$ and $v_{n}$ observed in experiments open a new window to the understanding of the collectivity phenomena in heavy-ion collisions.
However, it was also noticed that these measurements were based on 2-particle correlations, which might be suffered by non-flow effects, and they require sub-dividing such calculations and modeling resolutions associated with ESE due to finite event-wise multiplicities. Considering the computational constraints, this approach can not be performed easily in hydrodynamic calculations which usually are based on limited statistics compared to experimental data.

\subsection{Symmetric Cumulants ($SC$)}

A new type of observable for the analyses of flow harmonic correlations, \textit{Symmetric Cumulants} (originally named \textit{Standard Candles (SC)} in~\cite{Bilandzic:2013kga}), was proposed as $SC(m,n) = \left<\left<\cos(m\varphi_1\!+\!n\varphi_2\!-\!m\varphi_3-\!n\varphi_4)\right>\right>_{c}$.
If $m\neq n$, the isotropic part of the corresponding four-particle cumulant is given by:
\begin{eqnarray}
\left<\left<\cos(m\varphi_1\!+\!n\varphi_2\!-\!m\varphi_3-\!n\varphi_4)\right>\right>_c &=& \left<\left<\cos(m\varphi_1\!+\!n\varphi_2\!-\!m\varphi_3-\!n\varphi_4)\right>\right>\nonumber - \left<\left<\cos[m(\varphi_1\!-\!\varphi_2)]\right>\right>\left<\left<\cos[n(\varphi_1\!-\!\varphi_2)]\right>\right>\nonumber\\
&=&\left<v_{m}^2v_{n}^2\right>-\left<v_{m}^2\right>\left<v_{n}^2\right>.
\label{eq:4p_sc_cumulant}
\end{eqnarray}
For a detector with uniform acceptance in azimuthal direction, the asymmetric terms, e.g. $\left<\left<\cos(m\varphi_1\!-\!n\varphi_2)\right>\right>$, are averaged to zero.
The single event 4-particle correlation $\left<\left<\cos(m\varphi_1\!+\!n\varphi_2\!-\!m\varphi_3-\!n\varphi_4)\right>\right>$ could be calculated as: 
\begin{eqnarray}
 \left<\cos(m\varphi_1\!+\!n\varphi_2\!-\!m\varphi_3-\!n\varphi_4)\right>  &=&  \frac{1}{M(M-1)(M-2)(M-3)} \big[\left|V_{m}\right|^2\left|V_{n}\right|^2\!-\! 2\mathfrak{Re}\left[V_{m+n}V_{m}^*V_{n}^*\right]\!-\! 2\mathfrak{Re}\left[V_{m}V_{m-n}^*V_{n}^*\right]\nonumber\\
&+&\!\left|V_{m+n}\right|^2\!+\!\left|V_{m-n}\right|^2\!-\!(M\!-\!4)(\left|V_{m}\right|^2\!+\!\left|V_{n}\right|^2) +\!M(M\!-\!6)\big]\,.
 \label{eq:4p_sc_cumulant}
\end{eqnarray}
And the single event 2-particle correlation $\left<\left<\cos[m(\varphi_1\!-\!\varphi_2)]\right>\right>$ could be obtained as: 
\begin{equation}
\left<\cos[m(\varphi_1\!-\!\varphi_2)]\right>=\frac{1}{M(M-1)}\big[\left|V_{m}\right|^2\!-\!M\big]\,.
\label{eq:two_n_n}
\end{equation}
Then, the weights of $M(M-1)$ and $M(M-1)(M-2)(M-3)$ are used to get the event-averaged 2- and 4-particle correlations, as introduced in~\cite{Bilandzic:2013kga}.
Due to the definition, this new type of 4-particle cumulant $SC(m,n)$ is independent of the symmetry planes $\Psi_m$ and $\Psi_n$, and is expected to be less sensitive to non-flow correlations, which should be strongly suppressed in 4-particle cumulants. This was confirmed by the $SC(m,n)$ calculation using HIJING model~\cite{Wang:1991hta,Gyulassy:1994ew} which does not include anisotropic collectivity but e.g. azimuthal correlations due to jet production. It is observed that both $\left<\left<\cos(m\varphi_1\!+\!n\varphi_2\!-\!m\varphi_3-\!n\varphi_4)\right>\right>$ and $\left<\left<\cos[m(\varphi_1\!-\!\varphi_2)]\right>\right>\left<\left<\cos[n(\varphi_1\!-\!\varphi_2)]\right>\right>$ are non-zero, while $SC(m,n)$ are compatible with zero in HIJING simulations~\cite{ALICE:2016kpq}. This confirms that the $SC(m,n)$ measurements are nearly insensitive to non-flow correlations.
Therefore, it is believed that $SC(m,n)$ is nonzero if there is (anti-)correlations of $v_{n}$ and $v_{m}$. 
The investigation of $SC(m,n)$ will allow us to know whether finding $v_m$ larger than $\left<v_m\right>$ in an event will enhance or reduce the probability of finding $v_n$ larger than $\left<v_n\right>$ in that event, which provides a unique information for the event-by-event simulations of anisotropic flow harmonics.

\begin{figure}[h]
\centering
\includegraphics[width=0.49\textwidth]{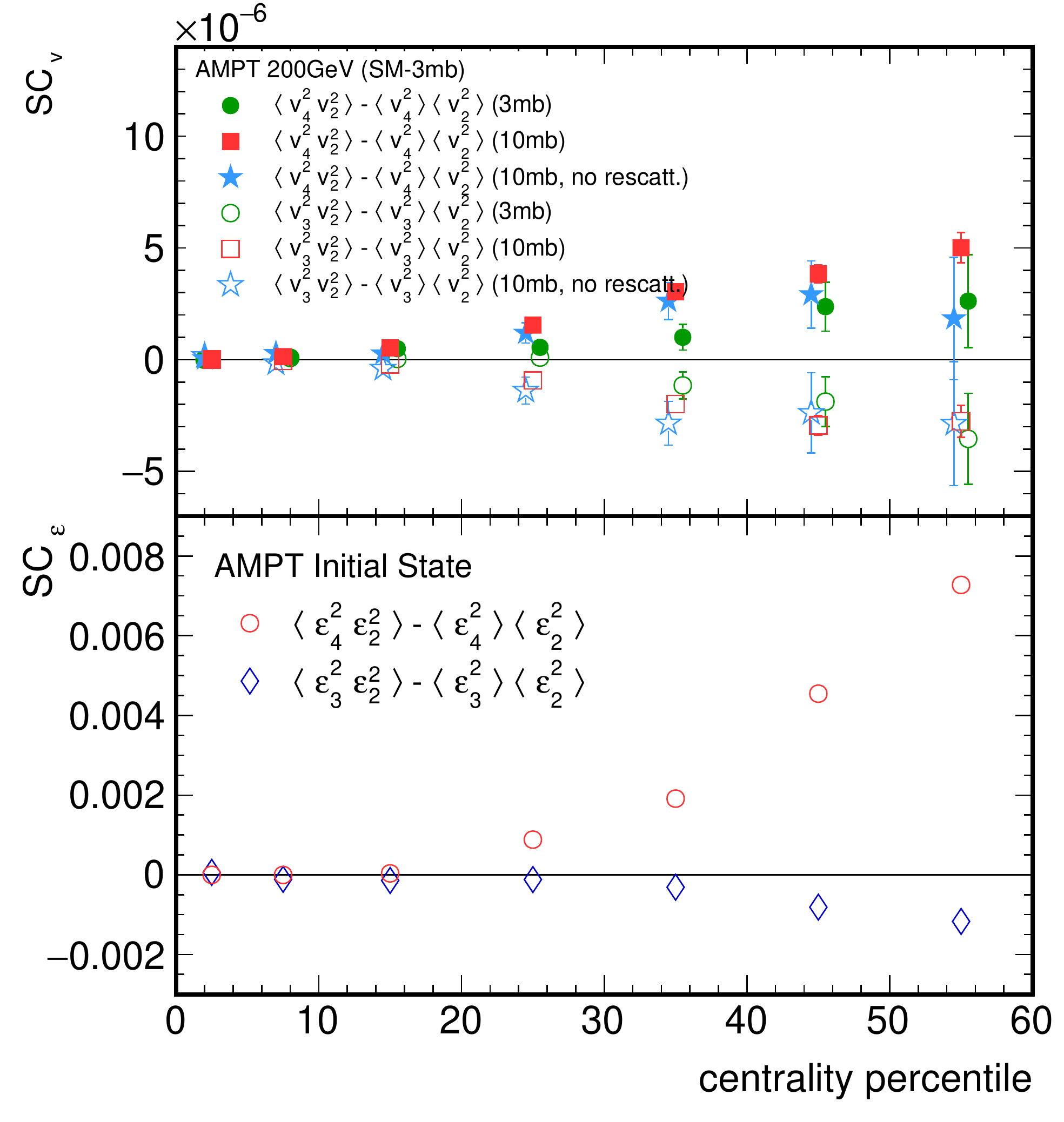}
\includegraphics[width=0.39\textwidth]{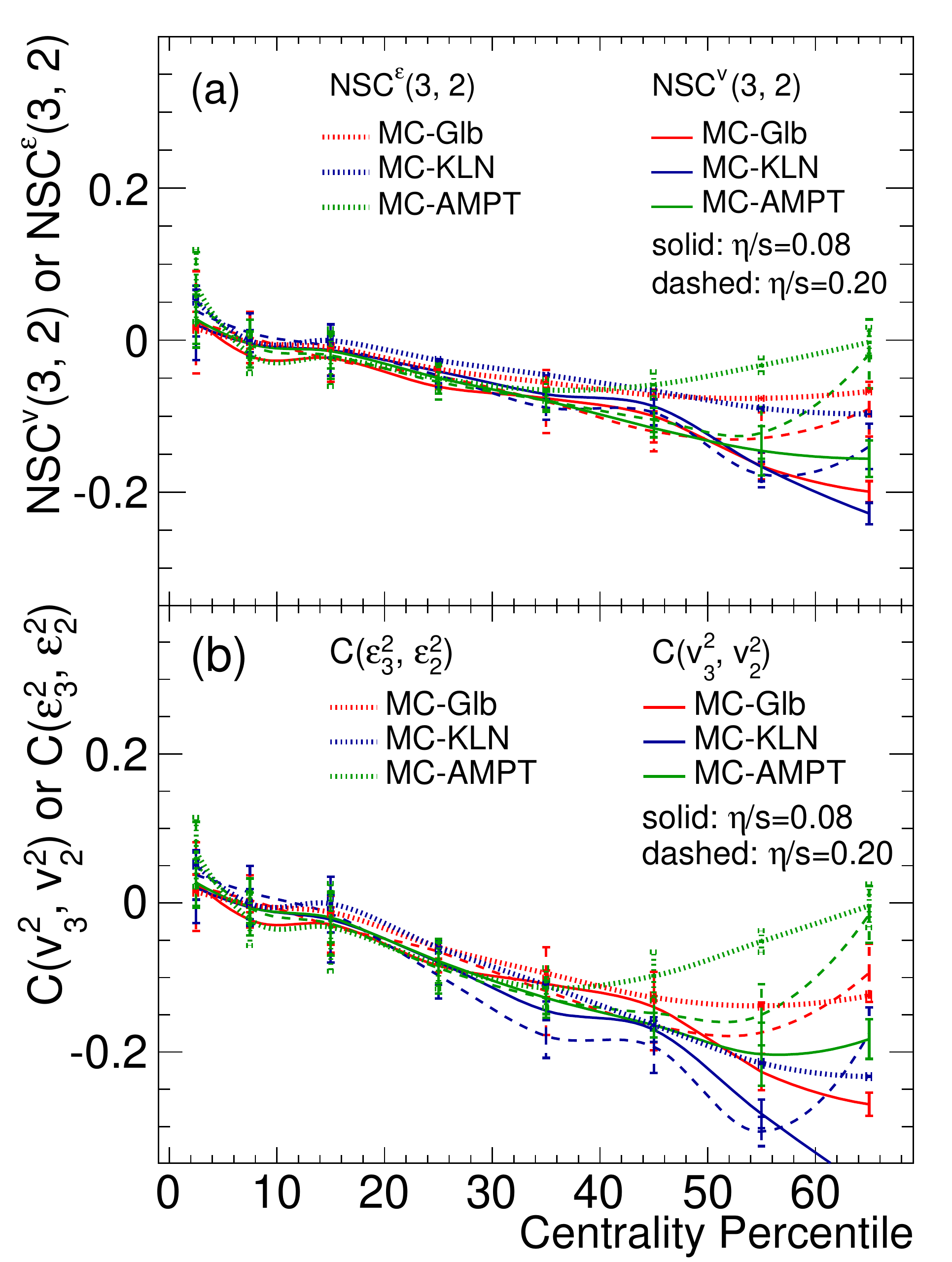}
\caption{(Color online) The centrality dependence of symmetric cumulants $SC(4,2)$ (red markers) and $SC(3,2)$ (blue markers) at $\sqrt{s_{_{\rm NN}}}$ = 2.76 TeV Pb--Pb collisions by ALICE. The AMPT calculations are presented by open markers. Figures taken from Ref.~\cite{Zhou:2015eya, Bilandzic:2013kga} (left) and~\cite{SC:hydro} (right).}
\label{fig:sc_AMPT_hydro}
\end{figure}

Figure~\ref{fig:sc_AMPT_hydro} shows the first calculation of $SC(4,2)$ (solid markers) and $SC(3,2)$ (open markers) as a function of centrality from AMPT model~\cite{Bilandzic:2013kga}. Non-zero values for both $SC(4,2)$ and $SC(3,2)$ are observed. The positive $SC(4,2)$ suggests a  correlation between the event-by-event fluctuations of $v_{2}$ and $v_{4}$, which indicates that finding $v_{2}$ larger than $\langle v_{2} \rangle$ in an event enhances the probability of finding $v_{4}$ larger than $\langle v_{4} \rangle$ in that event. On the other hand, the negative results of $SC(3,2)$ implies that finding $v_{2}$ larger than $\langle v_{2} \rangle$ enhances the probability of finding $v_{3}$ smaller than $\langle v_{3} \rangle$~\cite{Bilandzic:2013kga}. 

Several configurations of the AMPT model have been investigated to better understand the results based on AMPT simulations~\cite{Bilandzic:2013kga}. Partonic interactions can be tweaked by changing the partonic cross section: the default value is 10 mb, while using 3 mb generates weaker partonic interactions in ZPC~\cite{Zhou:2010us, Lin:2004en}. One can also change the hadronic interactions by controlling the termination time in ART. Setting NTMAX = 3, where NTMAX is a parameter which controls the number of time steps in ART (rescattering time), will effectively turn off the hadronic interactions~\cite{Zhou:2010us, Lin:2004en}. 
The $SC(4,2)$ and $SC(3,2)$ calculations for three different scenarios: (a) 3 mb; (b) 10 mb; (c)10 mb, no rescattering are presented in Fig.~\ref{fig:sc_AMPT_hydro} (left). It is found that when the partonic cross section is decreasing from 10 mb (lower shear viscosity) to 3 mb (higher shear viscosity), the strength of $SC(4,2)$ decreases. Additionally, the `10mb, no rescattering' setup seems to give slightly smaller magnitudes of $SC(4,2)$ and $SC(3,2)$. 

Further studies have been performed in AMPT initial conditions, based on the observable of $SC(m,n)_{\varepsilon}$ which is defined as $\left< \varepsilon_{m}^2 \varepsilon_{n}^2 \right> - \left< \varepsilon_{m}^2 \right> \left< \varepsilon_{n}^2 \right>$~\cite{Zhou:2015eya}. The centrality dependence of $SC(4,2)_{\varepsilon}$ and $SC(3,2)_{\varepsilon}$ are presented as red circles and blue diamonds in Fig.~\ref{fig:sc_AMPT_hydro} (left bottom). Positive and increasing trend from central to peripheral collisions has been observed for $SC(4,2)_{\varepsilon}$. In contrast, negative and decreasing trend was observed for $SC(3,2)_{\varepsilon}$ in the AMPT initial conditions. This shows that finding $\varepsilon_{2}$ larger than $\langle \varepsilon_{2} \rangle$ in an event enhances the probability of finding $\varepsilon_{4}$ larger than $\langle \varepsilon_{4} \rangle$, while in parallel enhancing the probability of finding $\varepsilon_{3}$ smaller than $\langle \varepsilon_{3} \rangle$ in that event. Same conclusions were obtained using MC-Glauber initial conditions~\cite{Zhou:2016fvj}.

Based on AMPT calculations, it seems that the signs of $SC(m,n)_{v}$ (for $m,n =$ 2, 3, 4) in the final state are determined by the correlations of $SC(m,n)_{\varepsilon}$ in the initial state, while its magnitude also depends on the properties of the created system. This clearly suggests that $SC(m,n)_{v}$ is a new promising observable to constrain the initial conditions and the transport properties of the system.

\begin{figure}[h]
\centering
\includegraphics[width=0.9\textwidth]{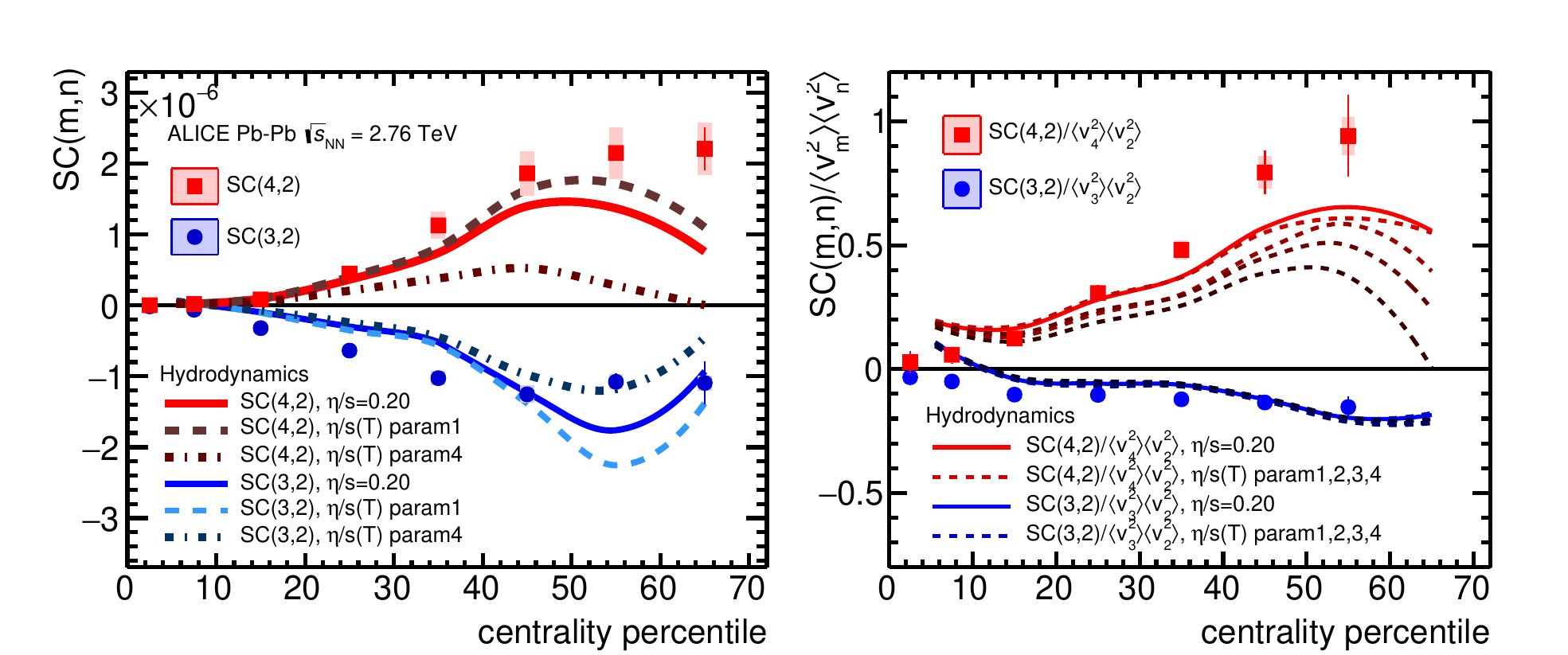}
\caption{(Color online) The centrality dependence of symmetric cumulants $SC(4,2)$ (red markers) and $SC(3,2)$ (blue markers) at $\sqrt{s_{_{\rm NN}}}$ = 2.76 TeV Pb--Pb collisions. Figures taken from Ref.~\cite{ALICE:2016kpq}. }
\label{fig:sc_ALICE}
\end{figure}

The first experimental measurements of centrality dependence of $SC(4,2)$ (red squares) and $SC(3,2)$ (blue circles) are presented in Fig.~\ref{fig:sc_ALICE} (left). Positive values of $SC(4,2)$ are observed for all centralities. This confirms a correlation between the event-by-event fluctuations of $v_{2}$ and $v_{4}$. On the other hand, the measured negative results of $SC(3,2)$ show the anti-correlation between $v_{2}$ and $v_{3}$ magnitudes.
The same measurements are performed using the like-sign technique, which is another powerful approach to estimate non-flow effects~\cite{Aamodt:2010pa}. It was found that the difference between correlations for like-sign and all charged combinations, which might be mainly due to non-flow effects, are much smaller compared to the magnitudes of $SC(m,n)$ itself. This further proves that non-zero values of $SC(m,n)$ measured in experiments cannot be explained by non-flow effects solely.

\begin{figure}[h]
\centering
\includegraphics[width=0.8\textwidth]{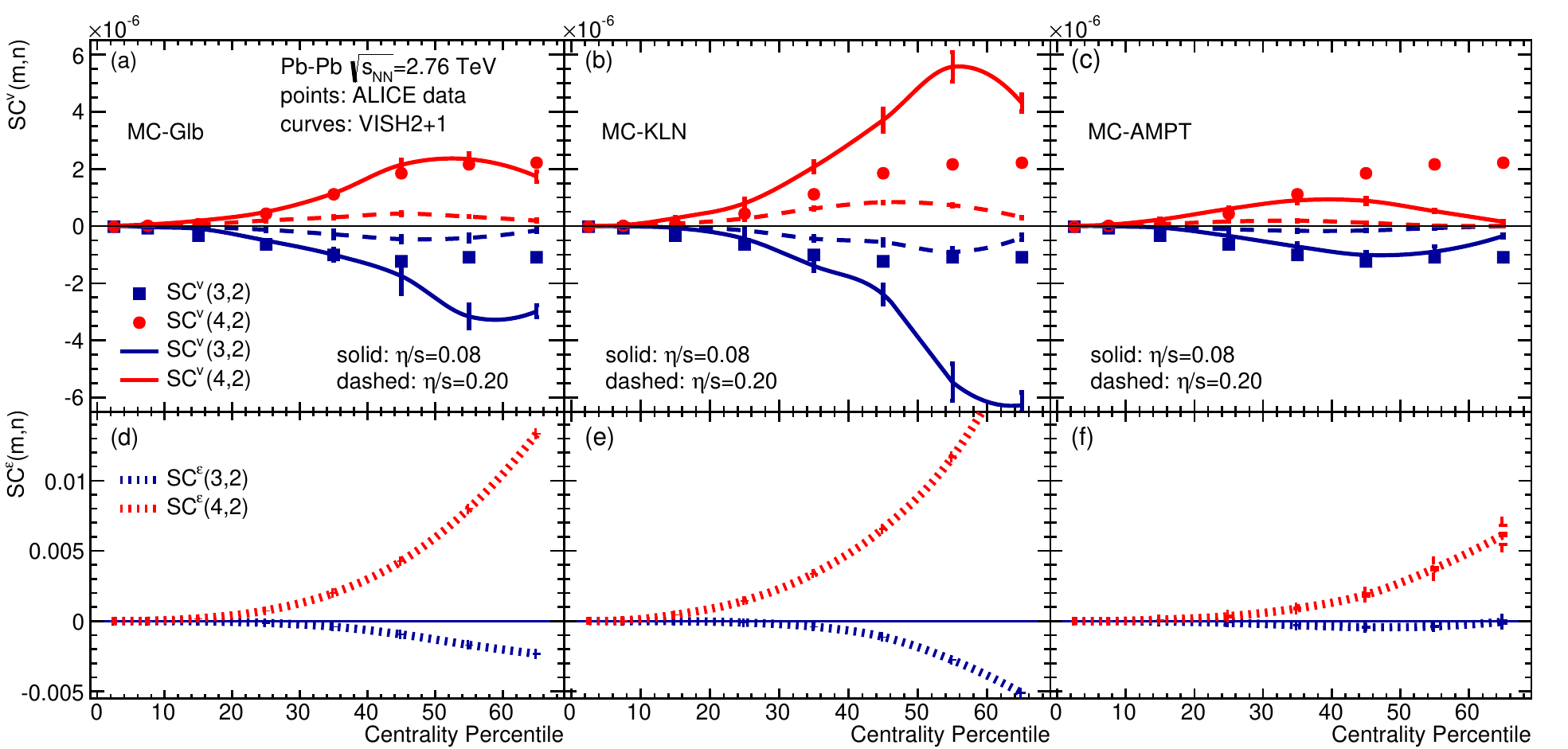}
\caption{(Color online) The centrality dependence of symmetric cumulants $SC(4,2)$ (red markers) and $SC(3,2)$ (blue markers) at $\sqrt{s_{_{\rm NN}}}$ = 2.76 TeV Pb--Pb collisions by VISH2+1 simulations. Figures taken from Ref.~\cite{SC:hydro}.}
\label{fig:sc_hydro}
\end{figure}
\begin{figure}[h]
\centering
\includegraphics[width=0.8\textwidth]{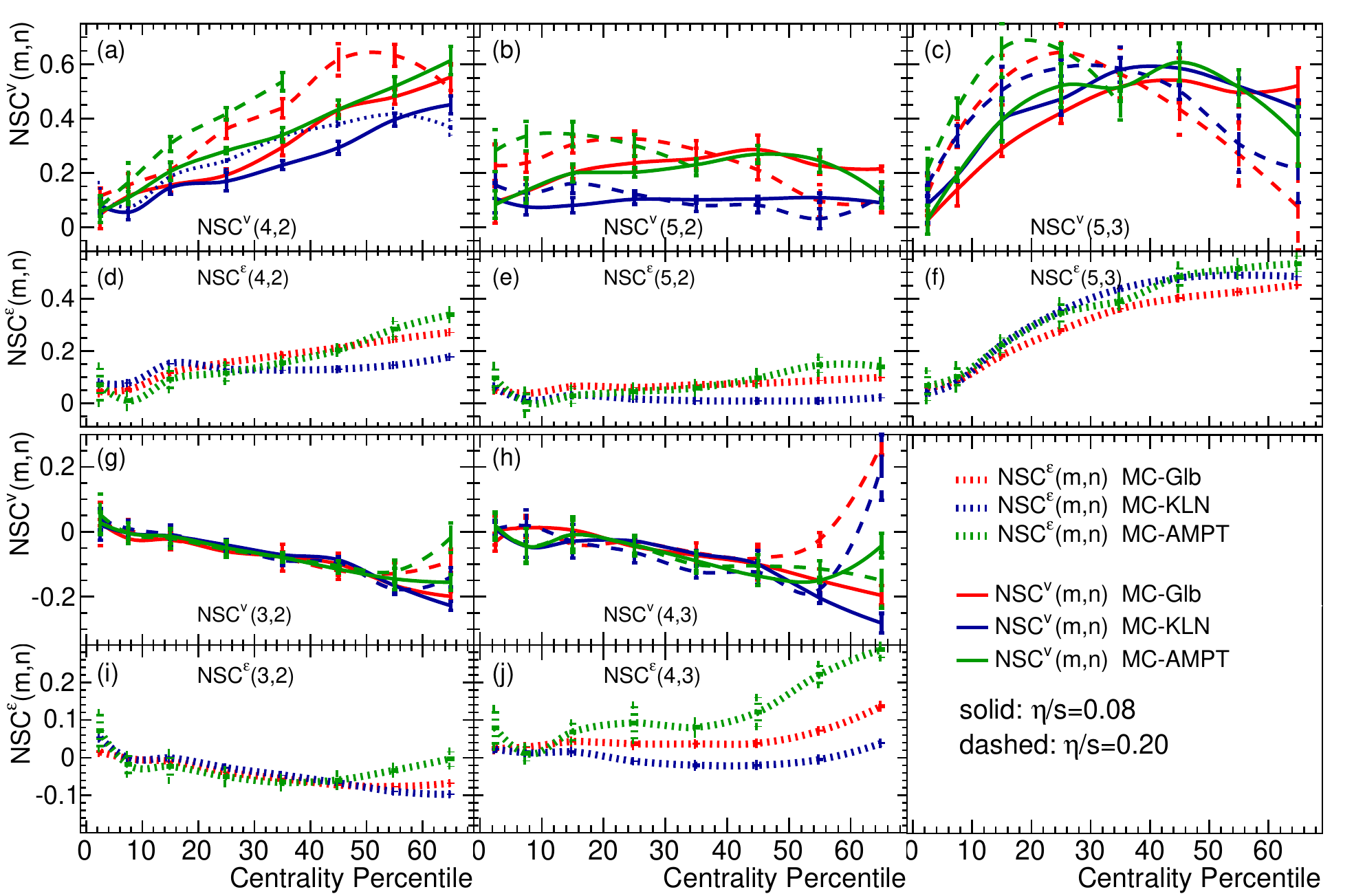}
\caption{(Color online) The centrality dependence of normalized symmetric cumulants ${\rm NSC}(m,n)$ at $\sqrt{s_{_{\rm NN}}}$ = 2.76 TeV Pb--Pb collisions by VISH2+1 simulations. Figures taken from Ref.~\cite{SC:hydro}.}
\label{fig:sc_hydro4}
\end{figure}

In addition, the comparison between experimental data and the event-by-event perturbative-QCD$+$saturation$+$hydro (``EKRT") calculations~\cite{Niemi:2015qia}, which incorporate both initial conditions and hydrodynamic evolution, is shown in Fig.~\ref{fig:sc_ALICE}. 
It was shown that this model can capture quantitatively the centrality dependence of individual $v_2$, $v_3$ and $v_4$ harmonics in central and mid-central collisions~\cite{Niemi:2015qia}. However, it can only qualitatively, but not quantitatively predict the $SC(m,n)$ measurements by ALICE. For a given $\eta/s(T)$ parameterization tuned by individual flow harmonic, the calculation can not describe $SC(4,2)$ and $SC(3,2)$ simultaneously for any single centrality. 
Experimental measurements are also compared to the VISH2$+$1 model calculations (see Fig.~\ref{fig:sc_hydro}), using various combinations of initial conditions (IC) from (a) MC-Glb; (b) MC-KLN and (c) MC-AMPT with $\eta/s=$ 0.08 and 0.20. It is noticed that the one with MC-Glb IC and $\eta/s=$ 0.08 is compatible with $SC(4,2)$ measurement and the calculation with MC-AMPT IC and $\eta/s=$ 0.08 can describe the $SC(3,2)$ measurement~\cite{SC:hydro}. However, just like EKRT calculations, none of these combinations is able to describe $SC(4,2)$ and $SC(3,2)$ simultaneously.
Thus, it is concluded that the new SC$(m,n)$ observables provide a better handle on the initial conditions and $\eta/s(T)$ than each of the individual harmonic measurement alone. 

After being presented for the first time at Quark Matter 2015 conference, preliminary results of $SC(4,2)$ and $SC(3,2)$ gained a lot of attention~\cite{Zhou:2015slf}. One of the key suggestions was to normalize $SC(m,n)$ by dividing with the products $\left<v_m^2\right>\left<v_n^2\right>$, in order to get rid of influences from individual flow harmonics. 
The results are shown in Fig.~\ref{fig:sc_ALICE}~(right), with normalized $SC(3,2)$ and $SC(4,2)$ observables by dividing with the products $\left<v_3^2\right>\left<v_2^2\right>$ and $\left<v_4^2\right>\left<v_2^2\right>$, respectively~\cite{ALICE:2016kpq}. 
The 2-particle correlations $\left<v_m^2\right>$ and $\left<v_n^2\right>$ are obtained with a pseudorapidity gap of $|\Delta\eta| > 1.0$ to suppress contributions from non-flow effects. 
It was shown in Fig.~\ref{fig:sc_hydro4} (top left) that the normalized $SC(4,2)$ observable exhibits a clear sensitivity to different $\eta/s$ parameterizations and the initial conditions, which provides a unique opportunity to discriminate between various possibilities of the detailed setting of $\eta/s(T)$ of the produced QGP and the initial conditions used in hydrodynamic calculations. On the other hand, normalized $SC(3,2)$ is independent of the setting of $\eta/s(T)$. In addition, it was demonstrated in Fig.~\ref{fig:sc_AMPT_hydro} (right) that the normalized $SC(3,2)$, also named $NSC^{v}(3,2)$ in the following text, is compatible with its corresponding observable $SC^{\varepsilon}(3,2)$ in the initial state. Thus, the $NSC^{v}(3,2)$ could be taken as golden observable to directly constrain initial conditions without demands for precise knowledge of transport properties of the system~\cite{SC:hydro}. 
Furthermore, none of existing theoretical calculations can reproduce the data, there is still a long way to go for the development of hydrodynamic calculations.

\begin{table}[htb]
  \begin{center}
    \footnotesize
    \begin{tabular*}{160mm}{@{\extracolsep{\fill}}ccc|cc}
      \hline
        \hline                                                                              
      \raisebox{0.01ex}{Observables}  &   Equations &   number of particles    & Exp.    &   Th. \\
       \hline
   $ \langle \langle \cos (2\varphi_{1} + 3\varphi_{2} - 2\varphi_{3} - 3\varphi_{4} ) \rangle \rangle_{c}$  &    $ \langle v_{2}^{2} \, v_{3}^{2} \rangle - \langle v_{2}^{2}  \rangle  \, \langle v_{3}^{2} \rangle$  &  4  & \cite{ALICE:2016kpq} & \cite{SC:hydro}, \cite{Bhalerao:2014xra}, \cite{Zhou:2015eya}\\
      $ \langle \langle \cos (2\varphi_{1} + 4\varphi_{2} - 2\varphi_{3} - 4\varphi_{4} ) \rangle \rangle_{c}$  &    $ \langle v_{2}^{2} \, v_{4}^{2} \rangle - \langle v_{2}^{2}  \rangle  \, \langle v_{4}^{2} \rangle$  &  4   & \cite{ALICE:2016kpq} & \cite{SC:hydro}, \cite{Bhalerao:2014xra}, \cite{Zhou:2015eya}, \cite{Giacalone:2016afq} \\
      $ \langle \langle \cos (2\varphi_{1} + 5\varphi_{2} - 2\varphi_{3} - 5\varphi_{4} ) \rangle \rangle_{c}$  &    $ \langle v_{2}^{2} \, v_{5}^{2} \rangle - \langle v_{2}^{2}  \rangle  \, \langle v_{5}^{2} \rangle$  &  4  &   & \cite{SC:hydro}, \cite{Bhalerao:2014xra}, \cite{Giacalone:2016afq}\\
      $ \langle \langle \cos (2\varphi_{1} + 6\varphi_{2} - 2\varphi_{3} - 6\varphi_{4} ) \rangle \rangle_{c}$  &    $ \langle v_{2}^{2} \, v_{6}^{2} \rangle - \langle v_{2}^{2}  \rangle  \, \langle v_{6}^{2} \rangle$  &  4 \\
       $ \langle \langle \cos (3\varphi_{1} + 4\varphi_{2} - 3\varphi_{3} - 4\varphi_{4} ) \rangle \rangle_{c}$  &    $ \langle v_{3}^{2} \, v_{4}^{2} \rangle - \langle v_{3}^{2}  \rangle  \, \langle v_{4}^{2} \rangle$  &  4 & & \cite{SC:hydro}\\
      $ \langle \langle \cos (3\varphi_{1} + 5\varphi_{2} - 3\varphi_{3} - 5\varphi_{4} ) \rangle \rangle_{c}$  &    $ \langle v_{3}^{2} \, v_{5}^{2} \rangle - \langle v_{3}^{2}  \rangle  \, \langle v_{5}^{2} \rangle$  &  4  &   & \cite{SC:hydro}, \cite{Bhalerao:2014xra}, \cite{Giacalone:2016afq}\\
      $ \langle \langle \cos (3\varphi_{1} + 6\varphi_{2} - 3\varphi_{3} - 6\varphi_{4} ) \rangle \rangle_{c}$  &    $ \langle v_{3}^{2} \, v_{6}^{2} \rangle - \langle v_{3}^{2}  \rangle  \, \langle v_{6}^{2} \rangle$  &  4 \\
      $ \langle \langle \cos (4\varphi_{1} + 5\varphi_{2} - 4\varphi_{3} - 5\varphi_{4} ) \rangle \rangle_{c}$  &    $ \langle v_{4}^{2} \, v_{5}^{2} \rangle - \langle v_{4}^{2}  \rangle  \, \langle v_{5}^{2} \rangle$  &  4 \\
      $ \langle \langle \cos (4\varphi_{1} + 6\varphi_{2} - 4\varphi_{3} - 6\varphi_{4} ) \rangle \rangle_{c}$  &    $ \langle v_{4}^{2} \, v_{6}^{2} \rangle - \langle v_{4}^{2}  \rangle  \, \langle v_{6}^{2} \rangle$  &  4 \\    
      $ \langle \langle \cos (5\varphi_{1} + 6\varphi_{2} - 5\varphi_{3} - 6\varphi_{4} ) \rangle \rangle_{c}$  &    $ \langle v_{5}^{2} \, v_{6}^{2} \rangle - \langle v_{5}^{2}  \rangle  \, \langle v_{6}^{2} \rangle$  &  4 \\
      ... & ... & 6 \\
  \hline   
     \hline
    \end{tabular*}
    \caption{\label{tab:AllSC} List of observables for correlations of flow harmonics, includes all combinations of symmetric 2-harmonics 4-particle cumulants (up to $v_{6}$). }    
  \end{center}
\end{table}

Predictions of relationship between other harmonics are provided in~\cite{SC:hydro} and shown in Fig.~\ref{fig:sc_hydro4}. Besides different sensitivities to IC and $\eta/s$ as seen above, the centrality dependence of the relationship between flow harmonics seems quite different. For instance, despite the differences in the initial conditions, a maximum value of $SC(5,3)$ is observed in central collision using $\eta/s=$ 0.20, while the maximum value is seen in more peripheral collision if $\eta/s=$ 0.08 is used.

Compared to the previous measurements of relationship between flow harmonics investigated using the ESE technique, the $SC(m,n)$ observable, provides a quantitative measure of these correlation strengths. 
Further investigations on relationship between flow harmonics using list of observables in Table~\ref{tab:AllSC} could be performed as a function of centrality, $p_{\rm T}$, $\eta$ $et.~al$, which is clearly non trivial.
Although one did not use the information of symmetry planes in both $ESE$ and $SC$ studies, recent study just reveals that flow harmonic correlations might be not completely independent on symmetry plane correlations~\cite{Giacalone:2016afq}. The proportionality relations between symmetric cumulants involving higher harmonics $v_{4}$ or $v_{5}$ and symmetry plane correlations is derived, which seems build the bridge between flow harmonic correlations and flow angle correlations (symmetry plane correlations). This might point out to a new direction of investigations of correlations between flow-vectors, and will shed a new light into the nature of fluctuating initial conditions and $\eta/s$ of the created QGP in heavy ion collisions.


\section{Summary}
In the past two decades, the underlying $p.d.f.$ of each single harmonic $P(v_{n})$ was investigated in great details. However, it is an open question at the moment how the joint underlying $p.d.f.$, including different order symmetry planes and harmonics, is described, especially if these correlations between different flow harmonics modify the single harmonics $P(v_{n})$. New observables discussed here begin to answer these open questions. Nevertheless, many more investigations between different flow harmonics, including higher order cumulants and higher harmonics, are necessary to reasonably constrain the joint $p.d.f$, and ultimately lead to new insights into the nature of fluctuation of the created matter in heavy ion collisions.
How to turn the multitude of measured and possibly measurable in future relationships between anisotropic flow harmonics into a focused search for correct initial conditions and detailed setting of $\eta/s$ is an exciting challenge for the theory community.

\section{Acknowledgments}
The author thanks J.J.~Gaardh\o je, K.~Gajdo\v{s}ov\'{a}, L.~Yan, J.Y.~Ollitrault and H.~Song for the comments on the manuscripts and fruitful discussions. 
The author is supported by the Danish Council for Independent Research, Natural Sciences, and the Danish National Research Foundation (Danmarks Grundforskningsfond).

%

%
\end{document}